\documentclass[runningheads]{llncs}
\usepackage{graphicx}
\usepackage{multirow}
\usepackage[subtle]{savetrees}

\begin{document}
%
%\title{Contribution Title\thanks{Supported by organization x.}}
%\title{Introducing the Korbit ITS}
%\title{Korbit: A Large-Scale, Open-Domain, Mixed-Interface Dialogue-Based ITS for STEM}
\title{A Large-Scale, Open-Domain, Mixed-Interface Dialogue-Based ITS for STEM}
%{\tt Korbit} ITS

%
%\titlerunning{Abbreviated paper title}
% If the paper title is too long for the running head, you can set
% an abbreviated paper title here
%
\author{}
\institute{}
%\author{Iulian Vlad Serban \and
%Varun Gupta \and
%Ekaterina Kochmar\inst{1} \and
%Dung D. Vu\inst{2,3} \and
%Robert Belfer\inst{3} \and
%Joelle Pineau \and
%Aaron Courville \and
%Laurent Charlin \and
%Yoshua Bengio
%}

\author{Iulian Vlad Serban\inst{1} \and Varun Gupta\inst{1} \and Ekaterina Kochmar\inst{1,2} \and \\ Dung D. Vu\inst{1,3} \and Robert Belfer\inst{1} \and  Joelle Pineau\inst{4,1} \and Aaron Courville \inst{4,1} \and \\ Laurent Charlin\inst{4,1} \and Yoshua Bengio\inst{4,1}
}

\authorrunning{Serban et al.}

\institute{
Korbit Technologies Inc., Montreal, Canada \\
\and University of Cambridge, Cambridge, United Kingdom \and
École de Technologie Supérieure, Montreal, Canada \and MILA (Quebec Artificial Intelligence Institute), Montreal, Canada}

% First names are abbreviated in the running head.
% If there are more than two authors, 'et al.' is used.
%
%\institute{Korbit Technologies Inc. Montreal, Quebec, Canada (\url{www.korbit.ai}) \and
%Cambridge University \and
%Mila -- Quebec Artificial Intelligence Institute
%}
%
\maketitle              % typeset the header of the contribution
%
%\vspace{-1.0cm}
\begin{abstract}
We present {\tt Korbit}, a large-scale, open-domain,
mixed-interface, dialogue-based intelligent tutoring system (ITS). 
{\tt Korbit} uses machine learning, natural language processing and reinforcement learning to provide interactive, personalized learning online.
{\tt Korbit} has been designed to easily scale to thousands of subjects, by automating, standardizing and simplifying the content creation process. Unlike other ITS, a teacher can develop new learning modules for {\tt Korbit} in a matter of hours. % and develop entire courses in a matter of days.
To facilitate learning across a wide range of STEM subjects, {\tt Korbit} uses a mixed-interface, which includes videos, interactive dialogue-based exercises, question-answering, conceptual diagrams, mathematical exercises and gamification elements.
{\tt Korbit} has been built to scale to millions of students, by utilizing a state-of-the-art cloud-based micro-service architecture.
{\tt Korbit} launched its first course in 2019 on machine learning, and since then over $7,000$ students have enrolled.
Although {\tt Korbit} was designed to be open-domain and highly scalable, A/B testing experiments with real-world students demonstrate that both student learning outcomes and student motivation are substantially improved compared to typical online courses. %traditional, video-based lectures and multiple choice quizzes.
%In this paper, we demonstrate that the intelligent tutoring system by utilizing deeply personalized hints leads to substantially improved student learning gains...
\end{abstract}
%The abstract should briefly summarize the contents of the paper in
%150--250 words.

%
%
%
\section{Introduction}
Intelligent tutoring systems (ITS) are computer programs powered by artificial intelligence (AI), which deliver real-time, personalized tutoring to students.
Traditional ITS implement or imitate the behavior and pedagogy of human tutors.
In particular, one type of ITS are dialogue-based tutors, which use natural language conversations to tutor students~\cite{nye2014autotutor}.
This process is sometimes called ``Socratic tutoring'', because of its similarity to Socratic dialogue~\cite{rose2001comparative}.
Newer ITS have started to interleave their dialogue with interactive media (e.g.\@ interactive videos and interactive web applets) -- a so-called ``mixed-interface system''.
It has been shown that ITS can be twice as effective at promoting learning compared to the previous generation of computer-based instruction and that ITS may be as effective as human tutors in general~\cite{kulik2016effectiveness}.

However, despite the fact that ITS have been around for decades and are known to be highly effective, their deployment in education and industry has been extremely limited~\cite{olney2018using,ritter2007cognitive}.
A major reason for this is the sheer cost of development~\cite{folsom2010plan,olney2018using}.
As observed by Olney~\cite{olney2018using}: \textit{``Unfortunately, ITS are extremely expensive to produce, with some groups estimating that it takes 100 hours of authoring time from AI experts, pedagogical experts, and domain experts to produce 1 hour of instruction.''}
For example, the creators of the ITS ``ITADS'' noted that their system took 26 months to develop~\cite{ramachandran2018itads}.
On the other hand, lower-cost educational approaches, such as massive open online courses (MOOCs), have flourished and now boast of having millions of learners around the world.
Indeed, it is estimated that today there are over 110 million learners around the world enrolled in MOOCs~\cite{shah2019moocstats}.
However, the learning outcomes resulting from learning in MOOCs depend critically on their teaching methodology and quality of content, and remains questionable in general~\cite{cavanaugh2015large,colvin2014learning,kirtman2009online,koedinger2015learning,koxvold2014moocs,otto2018s}.
In particular, recent research indicates that MOOCs having low levels of active learning, little feedback from instructors and peers, and few peer discussions tend to yield poor learning outcomes~\cite{koedinger2015learning,otto2018s}.
Furthermore, it is well-known that student retention in MOOCs is substantially worse than in traditional classroom learning~\cite{hone2016exploring}.
By combining the low cost and scalability of MOOCs with the personalization and effectiveness of ITS, we hope {\tt Korbit} may one day help to effectively teach and motivate millions of students around the world.

% , though possibly less effective,
%Learning in an introductory physics MOOC: All cohorts learn equally, including an on-campus class

\section{The Korbit ITS}

\begin{figure}
\vspace{-2.0em}
\includegraphics[width=\textwidth]{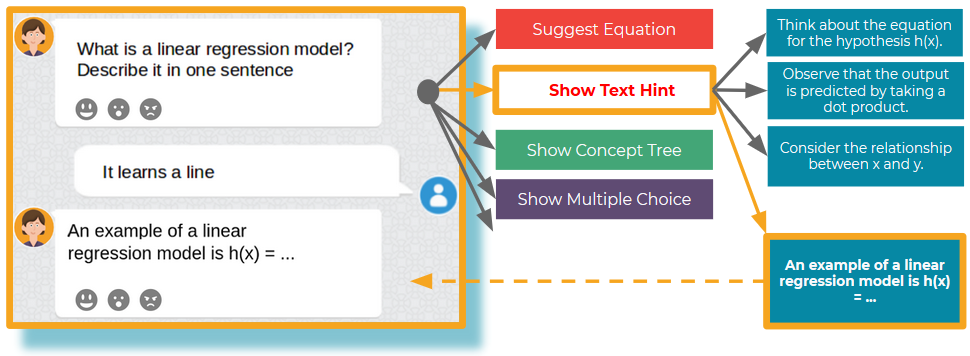}
\caption{The {\tt Korbit} ITS: An example dialogue illustrating how the ITS {\tt inner-loop} system selects the pedagogical intervention. The student gives an incorrect solution and afterwards receives a text hint.} \label{intervention_flow}
\end{figure}

%The platform is highly modular and scalable, and will soon be expanded with many more topics.

{\tt Korbit} is a large-scale, open-domain, mixed-interface, dialogue-based ITS, which uses machine learning, natural language processing (NLP) and reinforcement learning (RL) to provide interactive, personalized learning online.
The ITS has over 7,000 students enrolled from around the world, including students from educational institution partners and professionals from industry partners.
{\tt Korbit} is capable of teaching topics related to data science, machine learning, and artificial intelligence.
The platform is highly modular and, since it is easy to create new content, it will soon be expanded with many more topics.

Students enroll on the {\tt Korbit} website by selecting either a course or a set of skills they would like to study.
Students may also answer a few questions about their background knowledge.
Based on these, {\tt Korbit} generates a personalized curriculum for each student.
Following this, {\tt Korbit} tutors the student by alternating between short lecture videos and interactive problem-solving exercises.
The {\tt outer-loop} system decides on which lecture video or exercise to show next based on the personalized curriculum.
Currently, the ordering of videos and exercises is fully determined by the initial curriculum.
However, we are working on an extension to make the curriculum adapt during learning (for example, by adding new modules attacking student knowledge gaps on-the-fly).

During the exercise sessions, the {\tt inner-loop} system manages the interaction.
First, it shows the student a problem statement (e.g.,\@ a question).
The student may then attempt to solve the exercise, ask for help, or skip the exercise.
If the student attempts to solve the exercise, their solution attempt is compared against the expectation (i.e.\@ reference solution) using an NLP model.
If their solution is classified as incorrect, then the {\tt inner-loop} system will select one of a dozen different pedagogical interventions.
The pedagogical interventions include textual hints, mathematical hints, elaborations, explanations, concept tree diagrams, and multiple choice quiz answers.
The pedagogical intervention is chosen by an ensemble of machine learning models based on the student's profile and the last solution attempt.
Depending on the pedagogical intervention, the  {\tt inner-loop} system may either ask the student to retry the initial exercise or follow up on the intervention (e.g.,\@ with additional questions, confirmations, or prompts).
%The {\tt inner-loop} system is implemented as a finite-state machine.
%These models have been designed to provide the appropriate amount of scaffolding by choosing an intervention which is within the student's zone of proximal development (ZPD)~\cite{cazden1979peekaboo}.

%Include screenshots and explain %how the system works.

The {\tt Korbit} ITS is closely related to the line of work on dialogue-based ITS, such as the pioneering AutoTutor and the newer IBM Watson Tutor~\cite{ahn2018adaptive,graesser2005autotutor,graesser2001intelligent,nye2014autotutor,ventura2018preliminary}.
Although {\tt Korbit} is highly constrained compared to existing dialogue-based ITS, a major innovation of {\tt Korbit} lies in its modular, scalable design.
The {\tt inner-loop} system is implemented as a finite-state machine.
Each pedagogical intervention is a separate state, with its own logic, data and machine learning models.
Each state operates independently of the rest of the system, has access to all database content (including all exercises and lecture videos) and can autonomously improve as new data becomes available.
This ensures that the system gets better and better, that it can adapt to new content and that it can be extended with new pedagogical interventions.
Furthermore, the transitions between the states of the finite-state machine is decided by a reinforcement learning model, which itself is agnostic to the underlying implementation of each state and also continues to improve as more and more data becomes available.

%can be easily expanded with new types of pedagogical interventions.

%A major innovation introduced by  is the ability 

%\section{Related Work}
%1/2 page
%Talk about AutoTutor;
%Talk about IBM Watson Tutor;
%Talk about improving content authoring: AutoTutor works on having a simple authoring interface, Olney's work on automating it with %learners.

\section{System Evaluation}
We have conducted multiple studies to evaluate the {\tt Korbit} ITS.
Some of these studies have evaluated the entire system while others have focused on particular aspects or modules of the system.
Taken together, the studies demonstrate that the {\tt Korbit} ITS is an effective learning tool and that it overall improves student learning outcomes and motivation compared to alternative online learning approaches.

To keep things short, in this paper we limit ourselves and discuss only one of these studies. The study we present compares the entire system ({\tt Full ITS}) against an xMOOC-like system~\cite{daniel2012making}.
The purpose of this particular study is to evaluate 1) whether students prefer the {\tt Korbit} ITS or a regular MOOC, 2) whether the {\tt Korbit} ITS increases student motivation, and 3) which aspects of the {\tt Korbit} ITS students find most useful and least useful.
In an ideal world, {\tt Korbit} ITS would be compared against a regular xMOOC teaching students through lecture videos and multiple choice quizzes in a randomized controlled trial (a randomized A/B testing experiment).
However, it is not possible to compare against such a system in a randomized controlled trial, because it would create confusion and drastically offset our students expectations.\footnote{Indeed, we attempted this in an earlier study. However, during that study, as soon as students found out that they were assigned to the xMOOC system instead of the ITS system, they would complain to us, logout and create a new account to access the main ITS system.}
Therefore, in this study, we compare the {\tt Full ITS} against a reduced ITS, which appears identical to the {\tt Full ITS} and utilizes the same content (video lectures and exercise questions), but defaults to multiple choice quizzes 50\% of the time.
Thus, students assigned to the reduced ITS effectively spend about half of their interactions in an xMOOC-like setting.
We refer to this system as the {\tt xMOOC ITS}.

%\vspace{-0.5em}
\begin{table}[!t]
\centering
\caption{A/B testing results comparing the {\tt Full ITS} against the {\tt xMOOC ITS}: average time spent by students (in minutes), returning students (in \%), students who said they will refer others (in \%) and learning gain (in \%), with corresponding 95\% confidence intervals. The $^*$ and $^{**}$ shows statistical significance at 90\% and 95\% confidence respectively.}
\label{tab:a_b_testing_experiment}
\begin{tabular}{lllll}
System  & Time Spent & Returning Students & Will Refer Others & \hspace{0.1cm} Learning Gain \\
\hline
xMOOC ITS \hspace{0.1cm} & $22.98 \scriptsize{\pm 4.18}$ & $26.98\% \scriptsize{\pm 3.44\%}$ & $44.83\% \pm \scriptsize{9.00\%}$ & \multirow{2}{*}{$\mathbf{39.14\% \scriptsize{\pm 2.35\%}}$} \\
Full ITS & $\mathbf{39.86 \scriptsize{\pm 3.70}}^{**}$ & $\mathbf{31.69\% \scriptsize{\pm 1.92\%}}^{*}$ & $\mathbf{54.17\% \scriptsize{\pm 4.05\%}}$ & \\
%\hline
\end{tabular}
\vspace{-1.5em}
\end{table}

The experiment was conducted between October 7th and December 22nd, 2019, in an A/B testing setup with n=612 participants. Students who enrolled online were randomly assigned to either the {\tt Full ITS} (80\%) or {\tt xMOOC ITS} (20\%).
Students came from many different countries and were not subject to any selection or filtering process.
Apart from bug fixes and minor speed improvements, the system was kept fixed during this time period to limit confounding factors. After using the system for about 45 minutes, students were shown a questionnaire to evaluate the system.

% (n=126) 
%  (n=486)

Table \ref{tab:a_b_testing_experiment} shows the experimental results.
The average time spent in the {\tt Full ITS} was 39.86 min compared to 22.98 min in the {\tt xMOOC ITS}. As such, the {\tt Full ITS} yields a staggering 73.46\% increase in time spent.
In addition, the percentage of returning students and the percentage of students who said they would refer others to use the system is substantially higher for the {\tt Full ITS} compared to the {\tt xMOOC ITS}.
These results were also confirmed by the feedback provided by the students in the questionnaire.
Thus, we can conclude that students strongly prefer {\tt Korbit} ITS over xMOOCs and that the {\tt Korbit} ITS increases overall student motivation.

Table \ref{tab:a_b_testing_experiment} also shows the average student learning gain, which was observed to be 39.14\%. The learning gain is measured as the proportion of instances where a student provides a correct exercise solution after having receiving a pedagogical intervention from the {\tt Korbit} ITS. Thus, the pedagogical interventions appear to be effective.

Finally, in the questionnaire, 85.31\% of students reported that they found the chat equally or more fun compared to learning alone and 66.67\% of students reported that the chat helped them learn better sometimes, many times or all of the time.
%Apart from system bugs and Internet connectivity issues,
For the {\tt Full ITS}, 54.17\% of students reported that they would refer others to use {\tt Korbit} ITS.
In addition, students reported that the {\tt Korbit} ITS could be improved by more accurately identifying their solutions as being correct or incorrect and, in the case of incorrect solutions, by providing more relevant and personalized feedback.
\bibliographystyle{splncs04}
\bibliography{mybibliography}
%
%\begin{thebibliography}{8}
%\bibitem{ref_article1}
%Author, F.: Article title. Journal \textbf{2}(5), 99--110 (2016)

%\bibitem{ref_lncs1}
%Author, F., Author, S.: Title of a proceedings paper. In: Editor,
%F., Editor, S. (eds.) CONFERENCE 2016, LNCS, vol. 9999, pp. 1--13.
%Springer, Heidelberg (2016). \doi{10.10007/1234567890}

%\bibitem{ref_book1}
%Author, F., Author, S., Author, T.: Book title. 2nd edn. Publisher,
%Location (1999)

%\bibitem{ref_proc1}
%Author, A.-B.: Contribution title. In: 9th International Proceedings
%on Proceedings, pp. 1--2. Publisher, Location (2010)

%\bibitem{ref_url1}
%LNCS Homepage, \url{http://www.springer.com/lncs}. Last accessed 4
%Oct 2017
%\end{thebibliography}
\end{document}